# Critically assessing atavism, an evolution-centered and deterministic hypothesis on cancer

Bertrand Daignan-Fornier[1] and Thomas Pradeu[2,3]

[1] Univ. Bordeaux, CNRS, IBGC, UMR 5095, Bordeaux, France
[2] Univ. Bordeaux, CNRS, ImmunoConcEpT, UMR5164, Bordeaux, France
[3] Presidential Fellow, Chapman University, Orange, CA, USA

| Bertrand Daignan-Fornier: | b.daignan-fornier@ibgc.cnrs.fr |
|---|---|
| Thomas Pradeu: | thomas.pradeu@u-bordeaux.fr |

ORCID
BDF : 0000-0003-2352-9700
TP : 0000-0002-6590-3281

## ABSTRACT

Cancer is most commonly viewed as resulting from somatic mutations enhancing proliferation and invasion. Some hypotheses further propose that these new capacities reveal a breakdown of multicellularity allowing cancer cells to escape proliferation and cooperation control mechanisms that were implemented during evolution of multicellularity. Here we critically review one such hypothesis, named 'atavism', which puts forward the idea that cancer results from the re-expression of normally repressed genes forming a program, or toolbox, inherited from unicellular or simple multicellular ancestors. This hypothesis places cancer in an interesting evolutionary perspective that has not been widely explored and deserves attention. Thinking about cancer within an evolutionary framework, especially the major transitions to multicellularity, offers particularly promising perspectives. It is therefore of the utmost important to analyze why one approach that tries to achieve this aim, the atavism hypothesis, has not so far emerged as a major theory on cancer. We outline the features of the atavism hypothesis that , would benefit from clarification and, if possible, unification.

## INTRODUCTION

Because it lies among the most frequent causes of human death, cancer, as a disease, has received a lot of attention. It is quite surprising that, despite several millions of scientific articles published on cancer, there is not a consensual conceptual and theoretical framework in cancer biology. Even a universally accepted definition of cancer is lacking [1]. This is largely because cancer is a highly heterogeneous disease and cancers are often presented as a "family of diseases", despite the fact that most cancers share a set of common manifestations often summarized as uncontrolled proliferation, invasion, and metastases.

There is a strong need for well-articulated theoretical frameworks on cancer that would reveal some common underlying principles of cancer development and would make it possible to formulate novel testable predictions [2]. In the present paper, we explore one approach that presents itself as such a well-articulated, encompassing, and predictive theoretical framework, namely the 'atavistic theory'. But, before we explore this theory in detail, let us start with a short description of the current landscape of theoretical approaches to cancer.

There are many obstacles to building a general theory of cancer. One crucial obstacle is that, because of cancer heterogeneity, one counter-example is often not considered sufficient to falsify the hypothesis; therefore, the general plausibility of the hypothesis can be privileged over its refutability. There is an additional level of heterogeneity when considering hypotheses on cancer: some hypotheses deal with the distal and/or proximal causes of cancer (e.g. 'somatic mutation theory') while others concentrate on the modalities of its progression (e.g. 'tumors as organs' [3]). Some theories are centered on internal causes while others privilege external ones (e.g. 'viral origin' of cancer). Ideally, all hypotheses should account for the fact that cancer can result from external causes, such as infectious agents or mutagens, as well as endogenous causes, such as hormonal imbalance or tissue alteration and that less direct complex factors, such as diet or aging, further complicate the picture. Finally, when intending to compare cancer hypotheses, an additional layer of difficulty comes from the fact that, for each hypothesis, cell, tissue, or organ is proposed to be the major level of dysfunction, making it difficult to compare the different hypotheses in a single experimental model. Thus, several hypotheses explaining cancer have emerged over time and coexist in the literature, while their plausibility and soundness rely on very heterogeneous bases.

Probably the most influential theory on cancer is the somatic mutation theory (SMT). Historically, several viral infectious agents were identified as capable of generating cancer, initially suggesting that cancer could be the outcome of various infectious diseases, although in many cases no such agent could be identified. However, the later identification of *Helicobacter pylori* as an agent of stomach cancer has further substantiated the idea that we may still underestimate the frequency of cancers involving a viral or bacterial pathogen as a key causal agent. The current estimation of such infection-triggered cancers is in the 15-20% range [4]. Because many cancers are not thought to be associated with infectious agents, more proximal causes shared by all cancers were searched that would include infection-caused cancers but also cancers caused by radiations or hereditary predispositions, for example. This resulted in the SMT, which is today the prevalent hypothesis in cancer research. This theory says that cancer is a disease of genetic origin that initially affects a single cell and provides it with the capacity to proliferate under conditions where and when it should not. It stipulates that mutations in specific genes allow selection of hyper-proliferative cells that end-up invading adjacent tissues and propagating in the body in the form of metastases. A major strength of SMT is the fact that mutations, often (if not always), contribute to cancer and that several of these mutations have been clearly identified. The uncertainties and discussions concern more their precise causal role(s) in the process and how it is sequentially organized. How many mutations are necessary? How are they selected? In this theory, cancer is mostly genetic and accidental, although there are additional favoring circumstances that can be physiological or environmental. Importantly, onco-viruses and cancer-inducing somatic mutations often affect the same human proteins and contributed to identify key-player genes in the first place.

Despite the dominant position of SMT among cancer hypotheses, several intriguing observations that cannot be simply explained by SMT were reported [5], suggesting that mutations alone may not be sufficient to explain the whole picture of cancer development. For example, many chemical agents that are carcinogens are not mutagens and some of them, such as asbestos fibers, are thought to alter the tissues rather than the cells. This led to the emergence of an alternative theory, named tissue organization field theory (TOFT), stipulating that loss of tissue organization, and not genetic mutations, is at the origin cancer [5]. Basically, the idea is that *"…the control of cell division vanishes when the surrounded tissues lose their structure"* [6]. The TOFT extends a long tradition of 'field', tissue-level, explanations of cancer [7,8]. Whether SMT and TOFT are really incompatible or should be conciliated into a common hypothesis has been debated [6,9].

In addition to SMT and TOFT, there are hypotheses privileging the idea that cancer properties are not newly acquired, but instead *reemerge from the past*. In these hypotheses, cancer properties resurface from ontogeny ('embryonic origin') or from phylogeny ('atavism'). In these two cases, the mere arguments in favor of these theories are observed coincidence (e.g., gene expression) and theoretical plausibility, based on the idea that it is more likely that cancer develops using pre-existing properties, rather than creating new ones' *ex nihilo*. Basically, the embryonic and atavism hypotheses present cancer as the reactivation of functionalities that are repressed in the adult stage such as, for example, metabolic enzymes that are usually only expressed in the embryo and which expression is turned-off in the adult. Here we offer a systematic examination of the atavistic hypothesis, its roots, main claims, and arguments. We also discuss its connections to SMT and to the prevalent claim that cancer constitutes a breakdown of multicellularity [10]. We conclude by situating atavism into the broader context of theoretical oncology, especially as compared to other frameworks connecting evolutionary biology and cancer biology.

## THE ATAVISTIC HYPOTHESIS, AN EVOLUTION-CENTERED AND STRONGLY DETERMINISTIC VIEW OF CANCER

The atavistic hypothesis roots are ancient (see BOX 1), it clearly anchors cancer in long-term evolution (i.e., species evolution over billions of years), a viewpoint which is different from the 'within-organism evolution' view of SMT (see[11,12] for recent discussion). Specifically, the atavistic hypothesis opposes properties of unicellular ancestors, presumably reactivated during cancer, to those of complex multicellular organisms dealing with cancer. In this view, cancer cells would face stressful conditions reminiscent of those faced by ancestral unicellular cells in their environment [13–15]. These adverse conditions could tentatively be coped-with by reactivation of ancient functions, which would have remained repressed during evolution of multicellularity. Hence the atavistic hypothesis is centered on the idea of an incomplete transition to multicellularity [16], in which multicellular organisms would have conserved 'reactivatable' traces of their history, in the form of mechanisms previously selected to deal with it [15].

Demarcation from other hypotheses

Israel clearly places his hypothesis at a certain distance from the 'mutation-first' hypothesis:

> *The formidable survival efficiency conferred in every case by the genetic and phenotypic changes that accumulate in cancer cells during their malignant progression may be viewed as the result of combined and integrated genetic responses induced by damaging or threatening environmental events which are in themselves not necessarily mutagenic.* [13]

This idea was reaffirmed later by Davies and Agus, two other proponents of atavism:

> *The fact that cancer progresses in such an organized, systematic and predictable manner contradicts the notion of rogue defective cells running amok, and suggests*

> *instead a deeply-embedded pre-programmed repertoire of activity—a sort of 'cancer subroutine'— possibly triggered by mutations but not primarily driven by them.* [17]

These points are central to the atavistic hypothesis, the plausibility of which rests on the idea that the reactivation of an ancient (and normally repressed) program is much simpler and hence more likely than the expectation of multiple independent random mutations. This cost-effective point of view (less steps being more likely, by analogy to the use of parsimony in phylogeny) has been largely expressed by Mark Vincent [18]. The argument is that it would be much simpler (and hence more likely) to activate cancer properties collectively rather than individually through a process requiring multiple events (*random coincidence*) [18].

> *Cancer cells exhibit a characteristic suite of peculiar traits including ceaseless proliferation, genomic instability, immortality, insult resilience, paradoxical aerobic glycolysis, and host destruction. There is no a priori reason why these traits should consistently co-aggregate.* [18]

This is in direct opposition to the idea that cancer is the result of convergent within-organism evolution according to which similar conditions result in the selection of similar phenotypes.

> *The parsimonious explanation is not convergent evolution, but the release of an highly conserved survival program, honed by the exigencies of the Pre-Cambrian, to which the cancer cell seems better adapted; and which is recreated within, and at great cost to, its host.* [18]

> *… the fact that all cancer cells, in all people, at all times, come to exhibit this nearly identical set of core traits has to be seen as a fantastically coincident type of convergent evolution, although this is not often acknowledged.* [19]

In brief, parsimony is an important piece of argument in the atavistic hypothesis, and it operates at two different levels: first, in the atavistic hypothesis, there is no need to re-create pre-existing properties, they just need to be reactivated. Second, it seems much simpler to reactivate them collectively, via a common program, rather than individually and sequentially via random mutations. Of note, while parsimony is often used in phylogeny to compare the likelihood of evolution trajectories by comparing the number of successive steps needed in each path, it remains a probabilistic tool and in no case is a decisive instrument.

<u>Cancer as the result of the re-expression of a repressed program present in every somatic cell</u>
The pillar of the current forms of the atavistic hypothesis is the presumed existence of a repressed program (also referred to as an *ancient toolkit*) that is normally not expressed in human adult somatic cells. This strongly programmatic view of cancer differs significantly from the prevalent views that see cancer as a strongly accidental dysfunction, at either the cell (SMT) or tissue level (TOFT).

> *… cancer is an atavistic condition that occurs when genetic or epigenetic malfunction unlocks an ancient "toolkit" of preexisting adaptations, re-establishing the dominance of an earlier layer of genes that controlled loose-knit colonies of only partially differentiated cells, similar to tumors.* [20]

> *"Cancer viewed as a programmed, evolutionarily conserved life-form, rather than just a random series of disease-causing mutations, answers the rarely asked question of*

*what the cancer cell is for, provides meaning for its otherwise mysterious suite of attributes, and encourages a different type of thinking about treatment."* [18]

The key 'reason' alleged for such a program is that it would allow survival of the cell [13] or, as put by Vincent [18], cancer could be considered as a '*lifeboat*'. The view of cancer as a survival program, rather than a local dysfunction, emphasizes the apparent increased 'independency' of cancer cells toward the organism and are reminiscent of hypotheses favoring endogenous causes such as endo-parasitism or more generally the view of cancer cells as selfish or cheating actors [21,22]. Many properties associated with cancer are presumably part of the '*survival at any cost*' program, including the Warburg effect, genome instability, ceaseless proliferation (telomerase de-repression), immortality (apoptosis evasion) and insult resilience [18]. A comprehensive list of properties associated with cancer that could be of atavistic origin can be found in Table 16-2 of a book chapter by Vincent[23]. It is noteworthy that, by contrast to the SMT which is largely centered on cell proliferation, the atavistic hypothesis is focused on cell survival.

The mechanisms presumably triggering the reactivation of the 'repressed program' are not precisely defined. The authors generally do not exclude that mutations could play a role in this process, but other types of events are also envisioned.

> *However, even if mutations are, in a majority of cases, the first transforming events, it does not follow that all the ensuing genetic events are random or that they necessarily consist of mutations.* [13]
>
> *… damaging events, or events which are a threat to cell survival and function, may, whether mutagenic or not, induce a cascade of genetic responses which overcome the consequences of the initial damage and render the cell capable of dealing with several kinds of unrelated threats.* [13]
>
> *A useful analogy is a genie in a bottle, with the genie playing the role of the cancer program and the confining bottle representing the body's, as yet poorly understood, regulatory apparatus (including tumor suppression strategies). The bottle may be shattered in many ways (mutations caused by radiation damage, hypoxia, carcinogens, inflammation, stromal environmental changes, etc.) but, once released, the genie executes its agenda deterministically by accessing ancient, highly conserved, deeply entrenched and well-protected genetic pathways that control basic multicellular function like development, tissue regeneration and immune response.* [17]

Whatever the precise content of the program and how its execution is triggered, it is important to understand where this program stands and how it interferes with life in the absence of cancer. Basically, the program is based on specific gene expression and these genes are located in the genome. As put by Thomas and coworkers [24], the '*conserved survival program [is] encrypted in every eukaryotic cell*'. Genes contributing to the program are presumably in a latent form, suggesting that they are somehow turned-off. Of note, in one of the original formulations of the hypothesis, it is not ruled out that some of these genes could normally be active under specific conditions while others may have fallen into disuse.

> *Atavisms occur because genes for previously existing traits are often preserved in a genome but are switched off, or relegated to non-coding ("junk") segments of DNA.*
> *…*

> *Some of these pathways are still in active use in healthy organisms today, for example, during embryogenesis and woundhealing. Others have fallen into disuse, but remain, latent in the genome, awaiting reactivation.* [20]

The atavistic hypothesis is hence a gene-centered hypothesis, just as SMT, but, by contrast to SMT, atavism is not mutation-centered, nor selection-centered. Because the repressed program is deterministic, there is no need for selection to provoke tumorigenesis, it is all encrypted in the program itself. While there are several reasons why the atavistic hypothesis is appealing (see BOX 2), in the next two sections, we will list a series of evolution-related problems, as well as internal contradictions and divergences that appear to us as serious concerns regarding the atavistic hypothesis.

**EVOLUTION-RELATED PROBLEMS WITH THE ATAVISTIC HYPOTHESIS**

The framework underlying the atavism hypotheses is that evolution in general, and that of multicellularity in particular, proceeded by addition of layers of complexity that can take the form of new genes which are added to pre-existing genes and generate new capabilities.

> *In evading one layer of genetic regulation – turning proto-oncogenes into oncogenes – cancer mutations uncover a deeper, <u>older layer of genes</u> that code for behaviors that are often able to outsmart our best efforts to fight them.*
>
> *It took more than a billion years to evolve the eukaryotic genes present in Metazoa 1.0 and a further ~billion years to evolve the <u>sophisticated genetic and epigenetic overlay</u> that led to Metazoa 2.0.*
>
> *The genetic apparatus of the new Metazoa 2.0 was <u>overlain</u> on the old genetic apparatus of Metazoa 1.0.* [20] *(our emphasis)*

While this scenario of an increasing genetic complexity through evolution was once thought to be the most likely (Jacques Monod, for instance, in *Chance and necessity* [25], predicted one million proteins encoded by the human genome), whole genome sequencing has changed our view on genome complexity. Mechanisms of co-option as well as complex genetic interactions such as pleiotropy and epistasis, rather than simple addition of more genes, have emerged as a likely explanation for the low number of genes in complex organisms such as humans. Concerning multicellularity more specifically, several phylogenetic studies based on whole genome sequencing [26] rather argue for multicellularity developing (often by co-option) on a preset of protein domains already present in the unicellular ancestors and not by adding layers of complexity in the form of new 'multicellularity' genes. This was actually experimentally established, in the case of volvocines (green algae). Indeed, it was shown that expression of the Rb form of *Gonium pectorale* (a simple multicellular algae) in place of the endogenous form of Rb in its unicellular relative *Chlamydomonas reinhardtii*, was sufficient to make it colonial (simple multicellularity) [27].

Another major concern with the atavistic hypothesis is that the idea that 'disused genes' would be conserved in an intact functional form in the genome, waiting to be reactivated, is unlikely. Experimental evolution on myxobacteria allowed to establish that disused genes can be rapidly eliminated [28,29]. This was also illustrated by studies on the *Astyanax* cave fish eye [30]. These results suggested that, in absence of selective pressure or when there is a weak counter-selective pressure, disused traits can be repeatedly lost during evolution. Beside allowing survival of cancer cells, which does not seem to provide a clear evolutionary advantage, is there any selective pressure on the survival program to favor its maintenance for one billion years? This question is not directly addressed in the atavistic hypothesis while it severely affects its plausibility. Even weirder, these genes were proposed to be '*relegated to non-coding ("junk") segments of DNA*', as written by Davies and Lineweaver [20], a proposal seemingly suggesting that they might not be functional anymore. Where these genes lie in the genome and how they are maintained functional and silent over eons is critical for the hypothesis and not yet convincingly addressed neither conceptually nor experimentally. An alternative explanation could be that these genes are not disused. They could be only occasionally used (for example during embryonic development) but generally not used in the adult organism. Then, if these are genes required for the life cycle, what is 'atavistic here? Is it a gene network? A program? What does it mean at the molecular level? The repressed-gene hypothesis appears in contradiction with what we know on long-term evolution and gene conservation, while the repressed-program notion is so vague that the atavistic hypothesis loses its plausibility and testability. In brief, cell survival at any cost supposes the maintenance during billion-years of a program that is never successful, since cancer cell survival is ephemeral and does not give any obvious long-term advantage to the organism, which is precisely the entity selected in the million-year time scale separating the extant organism from its unicellular ancestor. Why should such an every-time-losing program be maintained during evolution? The program does not operate at the level at which natural selection operates, it should hence disappear rapidly, due to the lack of long-term selection, instead of being maintained in a repressed form as proposed in the atavistic hypothesis.

Of note, in the atavistic hypothesis proposed by Vincent [18], cancer cells have purposes and goals, giving the hypothesis a hint of finalism.

> *These organizing principles service <u>a single higher goal</u>: "any-cost cellular survivalism"*
>
> *Hence a broad spectrum, rapid response defense mechanism could have evolved, in which stochastic genomic reshuffling featured prominently, <u>providing a goal</u> to which everything else might be sacrificed.*
>
> *Escape and re-invention: The <u>purpose of cancer</u>*
>
> *The cancer cell's strange portfolio, including both aerobic glycolysis and some degree of (modulated) oxidative phosphorylation, along with hypermutation, aneuploidy, apoptotic disablement, immortality, ROS generation, DNA disrepair, asexual reproduction, differentiation block, and hyperproliferation, resurrects an ancient modus vivendi, <u>solely purposed</u> for opportunistic survival.*
>
> *If the cancer cell is a heuristic machine whose <u>sole purpose</u> is to invent its way to a future, then therapeutic refractoriness becomes more comprehensible*
>
> *Disturbingly, cancer exists as an encrypted potentiality, a proto-organism, in every eukaryotic cell, in every multicellular animal, including ourselves<u>: life's "Plan B", and purposed rather than accidental</u>.*[18] (<u>our emphasis</u>)

It is not clear to us whether this insistence on goals and purposes is merely a stylistic ease or a deliberate stance and whether the author appreciated how problematic this can be in a Darwinian perspective. At any rate, it somehow masks the implausibility of the hypothesis from a natural selection perspective. Importantly, this critique applies only to some, not all, proponents of atavism.

**INTERNAL CONTRADICTIONS AND DIVERGENCES, AMONG AUTHORS AND ALONG THE YEARS**

Because the atavism hypotheses were initially formulated in three distinct articles[13,18,20] and further diversified in subsequent work, they form a family of hypotheses with contradictions and divergences, that we think should be clarified by the proponents of the hypotheses. Examples of such divergences are discussed here.

<u>The ancestor</u>

In the various versions of the atavistic hypothesis, the ancestor, presumably at the origin of the repressed program that is reactivated upon cancer, is ill defined. Sometimes, the cancer survival program is compared to the well-known bacterial SOS response [13], which apparently has no strict equivalent in eukaryotes. In later articles, the evolutionary origin of the repressed program is much closer to human; for example, a proto-metazoan [20] or a *protozoan-like / type of unicellular holozoan opisthokont* [18], sometimes it is possibly a ciliate [31], as in the tentative explanation for a past foundation for genome instability (a highly frequent manifestation in cancer that is rather difficult to explain in an evolutionary perspective). The 'resuscitated ancestor' hence appears as a moving target, making any precise molecular prediction difficult. This malleability is obviously required for the *ad hoc* plausibility of each sub-hypothesis, but because the ancestral repressed program is at the very center of the atavistic hypothesis, the tension between the sub-hypotheses should be frontally discussed to clarify the conceptual framework. In the present forms of the hypothesis, it is not clear whether all the supporters of atavism have a common view on where in evolution the repressed program really comes from.

<u>Survival and proliferation</u>

Cancer is often defined as 'uncontrolled cell proliferation' and most hypotheses on cancer are centered on cell proliferation and how it is made possible, either by genetic means (SMT) or when surrounding tissues lose their structure (TOFT). By contrast, the atavistic hypothesis is centered on 'survival'. Indeed, the term 'survival' figures in the title of two founding papers [13,18] and is much more frequent in the texts (18 occurrences for *surviv\** in Israel 1996 [13]; 37 in Vincent 2012 [18]; 11 in Davies and Lineweaver 2011 [20]) than 'proliferation' (7 occurrences for *prolif\** in Israel 1996 [13]; 7 in Vincent 2012 [18]; 7 in Davies and Lineweaver 2011 [20]). This specificity of the atavistic hypothesis is a very important point, but the hypothesis lacks a clearly enunciated articulation with physiological and pathological cell proliferation. How does '*survival at any cost*' connect to hyper-proliferation? This is at odds with a possible unicellular origin of the 'repressed program' since in stressful environments, unicellular organisms most of the time suppress reproduction/proliferation to ensure long-term survival [10]. Does it mostly act by permitting proliferation once the cancer cell has been rescued from death by the 'repressed program'? Should survival here be understood as survival of the cell lineage rather than survival of the individual cells? This could make more sense but is not clearly stated by the authors.

The 'repressed program' is presumed to sit in every cell genome and its reactivation provides the cell with its cancer properties. However, nematodes that show no somatic proliferation at the adult stage (absence of somatic stem cells) show no cancer, suggesting that the repressed program never activates in nematodes or that its activation is not sufficient in the absence of somatic cell proliferation. If survival comes first and then allows proliferation (based on ancient unicellular properties), then the nematode should have cancer – except if it had lost the repressed program (together with, but independently from, somatic proliferation), which would be a very unlikely coincidence. The plausibility of the current atavistic hypothesis as an auto-sufficient cause of cancer is severely challenged by this apparent dependency on prior proliferation.

The SMT attraction

In its initial forms, the atavistic hypothesis differed significantly from the SMT in two important ways. First, in the SMT, cancer is understood as a multistep disease, while the 'repressed program' proposed by atavism makes it understandable using a single step (reactivation). Second, Darwinian selection, in the SMT, is critical for the succession of steps to occur and eventually result in cancer by a convergent evolution type of mechanism. By contrast, in the atavistic hypothesis, once the program is launched, there is no need for Darwinian selection to express all the properties of cancer cells, since cancer properties in the atavistic hypothesis are just passive outcomes of the repressed program. Although both atavism and SMT are gene-centered hypotheses, they are very different and largely incompatible.

> *The existence of such a toolkit implies that the progress of the neoplasm in the host organism differs distinctively from normal Darwinian evolution.* [20]

These two major differences have been largely wiped out in two recent publications, drifting the atavistic hypothesis continuously closer to the SMT. First, the unique-program initial idea is challenged by the fact that traits such as genome instability or the Warburg effect are clearly frequent but not universal in cancer. This is at odds with a deterministic program leading to a unique outcome that was initially proposed in the *genie-in-a-bottle* metaphor: "…*once released, the genie executes its agenda deterministically...*" [17]. In a more recent publication entitled "*Cancer progression as a sequence of atavistic reversions*" [32], the authors abandon the idea of the unique-program (the word 'program' does not even appear anymore in the paper) and shift toward a series of sequential reversions that phenotypically resemble ancestral behaviors and are thus called 'atavistic' but are not said anymore to proceed from the reactivation of a preexisting program. The hypothesis here shifts from global resuscitation of an ancestral logic to sequential reversion of some multicellular traits and simultaneously the notion of 'program' gives way to the notion of 'modalities'. Finally, the notion of disuse is abandoned, and nothing seems 'repressed' anymore. As expected in such an attraction toward SMT, proliferation is much more used than survival in the 2021 paper (15 times *vs* 4 times).

> *Here, we propose that cancer onset and progression involve more than a one-off multicellular-to-unicellular reversion, and are better described as a series of reversionary transitions.*
>
> *We argue that cancer is not a single atavism, but a series of atavisms.*
>
> *Significantly, cancer does not evolve the hallmark properties ab initio; rather, neoplastic phenotypes are preexisting modalities latent in the genome, retained because they play critical roles in key processes such as embryogenesis, tissue maintenance and wound healing.* [32]

This renewed hypothesis has its own consistency, but it totally loses two major aspects of the initial hypothesis, namely the plausibility associated with parsimony (claiming that one step is sufficient to explain all the cancer properties, a situation that would be difficult to explain based on random appearance) and the deterministic aspect associated with the programmatic viewpoint.

> *Unlike familiar morphological atavisms, such as supernumerary nipples which involve a one-off ontogenic transition cancer is a multistage process in the direction of increasing malignancy.* [32]

Then how come that a whole set of properties recurrently co-appear in cancer? The *random coincidence* objection made toward SMT [18] clearly backfires this new hypothesis and, accordingly, in a parallel work, the possibility of some convergent evolution was reconsidered by Vincent and coworkers [24] in an article entitled "*Cancer adaptations: Atavism, de novo selection, or something in between?*". These two more recent publications [24,32] strongly weaken the initial claims and tend to reposition the atavistic hypothesis very close to the SMT, which strongly contrasts with atavism's initial intentions. What important differences might remain? Based on the work published by Lineweaver and Davies[31], a main difference between SMT and the atavistic hypothesis is the time scale of evolution considered to account for cancer occurrence. It counts in years for "within-organism evolution" (SMT) and billions of years in the atavism hypothesis. The second major difference reside in the role played by selection and arises from the first difference mentioned above. Within organism selection potentially operates by all possible means (mutations, overexpression, chimeras…) affecting possibly both preexisting or new traits to increase fitness, while in the atavism hypothesis, only *loss-of-new-functions* and *gain-of-old-functions* are altered. Whether these differences can be reconciled or not will probably be a matter for future debates.

One specificity of the atavistic hypothesis is that it is highly deterministic. The initial trigger signal might result from an 'accident' but the following steps are highly predictable because they are presumed 'prewritten'. When formulating hypotheses about cancer, there is a constant tension between chaos and predictability. The chaotic part of cancer is largely due to genome instability, which is found at high frequency in tumors. On the other hand, some sequential recurrence is observed in the tumorigenesis process as well as some level of organization in the tumor outcome (e.g., tumor as an organ). Such regularities are not understood as produced by chaos. Hence, hypotheses on cancer have to combine chaos and predictability. The atavistic hypothesis is highly deterministic with its constant appeal to a 'program', and *de facto* there are multiple computer analogies in Davies and Lineweaver's work [20]. The notion of program is thought here as a 'pre-written' sequence (cascade) of events with a pre-known 'long-term' outcome (suggesting a purpose). The view of life as the execution of pre-written programs (metabolic, cell cycle, cell death, development…), has been widely questioned by biologists and philosophers alike (e.g. [33,34]). It is notable that in the more recent articles about the atavistic hypothesis the notion of program is disappearing – it is not even mentioned in the 2021 paper by Lineweaver colleagues [32]. This leaves us with serious doubts on what subsists of the initial hypothesis formulated by the same authors in 2011 [20]. Indeed, at the end, what is left of the idea of a unique program taking place independently of selection? In 2024, what is left of the atavistic hypothesis beside the vague idea that cancer could somehow be a breakdown of multicellularity?

**CONCLUSION**

Despite the difficulties of atavism, this approach may prove to have been useful, in retrospect, from at least two perspectives. First, it constitutes one of the few attempts to build an overarching theory of cancer, in a field where broad-ranging theories have been scarce. Second, it has participated in the emergence of more specific conceptual and theoretical discussions connecting cancer biology with other fields, especially evolutionary biology and research on multicellularity. Indeed, there is a growing and interesting literature on cancer and multicellularity [10,22], some of which alludes to atavism. The present critique of atavism does not imply at all a critique of this stimulating literature, in which recent and less extreme propositions on how cancer could be connected to loss of multicellularity stand out. For instance, in her 2020 paper, Aurora Nedelcu proposes two alternatives [10]. Cancer could indirectly alter sophisticated cell cooperation properties that were acquired during evolution of multicellularity or more directly impair specific mechanisms controlling these properties. She draws many potentially fruitful consequences of these alternatives from an experimental and even therapeutic viewpoint.

Should the initial atavism hypothesis be experimentally tested, it would be necessary to identify key genes of the 'repressed program', then show that their absence abolishes cancer (or at least some specific cancer properties) and establish whether these genes are expressed during normal life cycle or extinct all the time except during cancer (if such repressed genes were specifically reactivated during tumorigenesis, then they should be easy to find in the terabytes of cancer transcriptome data publicly available).

Because cancer is ill defined and highly heterogeneous, it has been very difficult to build-up a satisfying conceptual framework of cancer development. Exploring the various hypotheses available in the literature allows to narrow down the likelihood of specific explanations and eliminate the less plausible hypotheses. While the atavistic hypothesis was initially built-up on plausibility associated with parsimony, in more recent years it lost this asset and appears more and more as an evolution-centered byproduct of SMT. Rethinking cancer in an evolutionary perspective is interesting and could be fruitful but, in our opinion, should be done away from the poorly founded 'program' standpoint that is at the center of the atavistic hypothesis.

## ACKNOWLEDGMENTS

The authors thank Aurora Nedelcu, Jean Clairambault and Samir Okasha for their comments on the manuscript. This work was supported by the NewMoon research program of the University of Bordeaux. The research of TP is funded by the Gordon and Betty Moore Foundation through grant GBMF9021.

Data availability statement: Data sharing is not applicable to this article as no new data were created or analyzed in this study.